%% file: main.tex
\documentclass[10pt, conference, letterpaper]{ IEEEtran}

\IEEEoverridecommandlockouts

\renewcommand{\baselinestretch}{0.97}
\usepackage{cite}
\usepackage{amsmath,amssymb,amsfonts, amsbsy}
\usepackage{graphicx}
\usepackage{pifont}
\usepackage{textcomp}
\usepackage{xcolor}
\usepackage{comment}
\usepackage{pifont}
\usepackage{multirow}
\usepackage{varwidth}
\usepackage{url}
\usepackage{soul}
\usepackage{enumitem}
\usepackage{graphicx}
\usepackage{capt-of}
\usepackage{booktabs} 
\usepackage{array} 
\usepackage{verbatim} 
\usepackage{enumitem}
\usepackage[hidelinks]{hyperref}

\usepackage{bbm}
\usepackage{caption}

\DeclareMathOperator*{\argmin}{argmin} 
 
\usepackage{varioref}
\usepackage{cleveref}
\usepackage{xltabular}
\usepackage{listings}
\usepackage{xcolor}
\usepackage{paralist}

\usepackage{amsthm}

\usepackage{mathrsfs}
\usepackage{bbm}

\usepackage{bm}
\usepackage{mdframed}

\usepackage[ruled,linesnumbered,noend]{algorithm2e}

\usepackage{algorithmicx}
\usepackage{algpseudocode}

\def\BibTeX{{\rm B\kern-.05em{\sc i\kern-.025em b}\kern-.08em
    T\kern-.1667em\lower.7ex\hbox{E}\kern-.125emX}}
\begin{document}

 
 \newcommand{\twodots}{\mathinner {\ldotp \ldotp}}

\makeatletter
\def\thm@space@setup{\thm@preskip=4pt
\thm@postskip=3pt}
\makeatother

\title{Workflow Optimization for Parallel Split Learning}

\author{\IEEEauthorblockN{Joana Tirana$^{\star}$, Dimitra Tsigkari$^{\dagger}$, George Iosifidis$^{\dagger}$, Dimitris Chatzopoulos$^{\star}$}
\IEEEauthorblockA{$^{\star}$School of Computer Science, University College Dublin \& VistaMilk SFI, Ireland\\
$^{\dagger}$Delft University of Technology, The Netherlands\\
joana.tirana@ucdconnect.ie, D.Tsigkari@tudelft.nl, G.Iosifidis@tudelft.nl, dimitris.chatzopoulos@ucd.ie} \thanks{ This manuscript will appear in \textbf{Proceedings of IEEE INFOCOM 2024}. 
Acknowledgments: This work has been supported by  SFI-VistaMilk grant no 16/RC/3835 and  EU Horizon project no 101092912~(MLSysOps). }}

\maketitle

\begin{abstract}
Split learning (SL) has been recently proposed as a way to enable resource-constrained devices to train multi-parameter neural networks (NNs) and participate in federated learning (FL). In a nutshell, SL splits the NN model into parts, and allows clients (devices) to offload the largest part as a processing task to a computationally powerful helper. In parallel SL, multiple helpers can process model parts of one or more clients, thus, considerably reducing the maximum training time over all clients (makespan). In this paper, we focus on orchestrating the workflow of this operation, which is critical in highly heterogeneous systems, as our experiments show. In particular, we formulate the joint problem of client-helper assignments and scheduling decisions with the goal of minimizing the training makespan, and we prove that it is NP-hard. We propose a solution method based on the decomposition of the problem by leveraging its inherent symmetry, and a second one that is fully scalable. A wealth of numerical evaluations using our testbed’s measurements allow us to build a solution strategy comprising these methods. Moreover, we show that this strategy finds a near-optimal solution, and achieves a shorter makespan than the baseline scheme by up to 52.3\%.
\end{abstract}

\input{sections/intro}
\input{sections/related}
\input{sections/model}
\input{sections/problem}
\input{sections/evaluation}
\input{sections/conclusion}

\clearpage
\bibliographystyle{IEEEtran}
\bibliography{IEEEabrv,references}

\end{document}

%% file: sections/intro.tex
\section{Introduction}
\label{sec:intro}

\textbf{Motivation.} The proliferation of devices that collect voluminous data through their sensors motivated the design of client-based distributed machine learning~(ML) protocols, like federated learning~(FL)~\cite{mcmahan2017communication}. In FL, the training process is organized in training rounds that include local model updates at the devices (that act as \emph{clients}) and the aggregation of all the clients’ models at a server~(the \emph{aggregator}). During this process,   clients keep their dataset locally, while only sharing their model's updates with the aggregator.

Some of the main challenges in FL are: 1)~system heterogeneity; 2)~communication overhead;
 3)~constrained resources, i.e., clients of limited memory and computing capacities~\cite{lim2020federated}.
As a result of these factors, the training time of some clients might be prohibitively long, thus, slowing down this cross-silo distributed ML process.
Indeed, these clients~(stragglers) increase    the \emph{training makespan}, i.e., the maximum 
training time over all clients, which is a key metric in highly heterogeneous systems because of the synchronous nature of FL~\cite{lim2020federated, pilla2021optimal}.
 While state-of-the-art FL approaches
propose ways of alleviating this phenomenon, e.g., via model pruning~\cite{jiang2022model} or asynchronous protocols~\cite{sprague2018asynchronous},  they may compromise the accuracy of the produced model. Moreover,  clients with limited memory capacity (e.g., IoT devices) might not be able to participate in FL  processes that train large ML models.

Split learning (SL) protocols have been recently proposed in order to enable resource-constrained clients to train neural networks (NNs) of millions of parameters~\cite{vepakomma2018split}. 
In SL,  clients offload a part of their training task to a \emph{helper}, which could be a Virtual Machine~(VM) on the cloud or a lightweight container in a base station in beyond 5G networks. 
Formally, a NN comprising $L$ layers\footnote{Throughout this manuscript, a ``layer" is the NN's model part that is indivisible, i.e., it cannot be further partitioned into more layers.} $(1,\ldots,L)$ is split into three parts (part-1, part-2, and part-3) of consecutive layers $([1,\ldots, \sigma_1]$, $[\sigma_1+1,\ldots,\sigma_2]$, $[\sigma_2+1,\ldots,L])$ using 2 \emph{cut layers} 
$\sigma_1$~and~$\sigma_2$. 
Then, part-1 and part-3 are processed at the clients, and part-2 at the helper. This allows the resource-constrained clients to offload computationally demanding processes to the helper.

In conventional SL, the clients share the same part-2, and the helper collaborates with each client in a sequential order to train the model parts. 
This can lead to long delays in the training process depending on the number of participating clients.
Whereas, in parallel SL~\cite{thapa2022splitfed,jeon2020privacy}, the helper allocates a different version of part-2 for each client, allowing clients to make parallel model updates.  At the end of each training round, all clients synchronize their model versions, and thus, the training makespan remains a key metric.
Essentially, parallel SL is the integration of SL in the FL protocol.
In fact, parallel SL reduces the makespan~(when compared to the conventional SL)  without compromising the model accuracy~\cite{thapa2022splitfed,zhang2023privacy, 10.1145/3446382.3448362}.

\begin{figure}[t]
	\centering  
	\includegraphics[width=8.9cm, trim={1.41cm 9.94cm 3.79cm 0.575cm},clip]{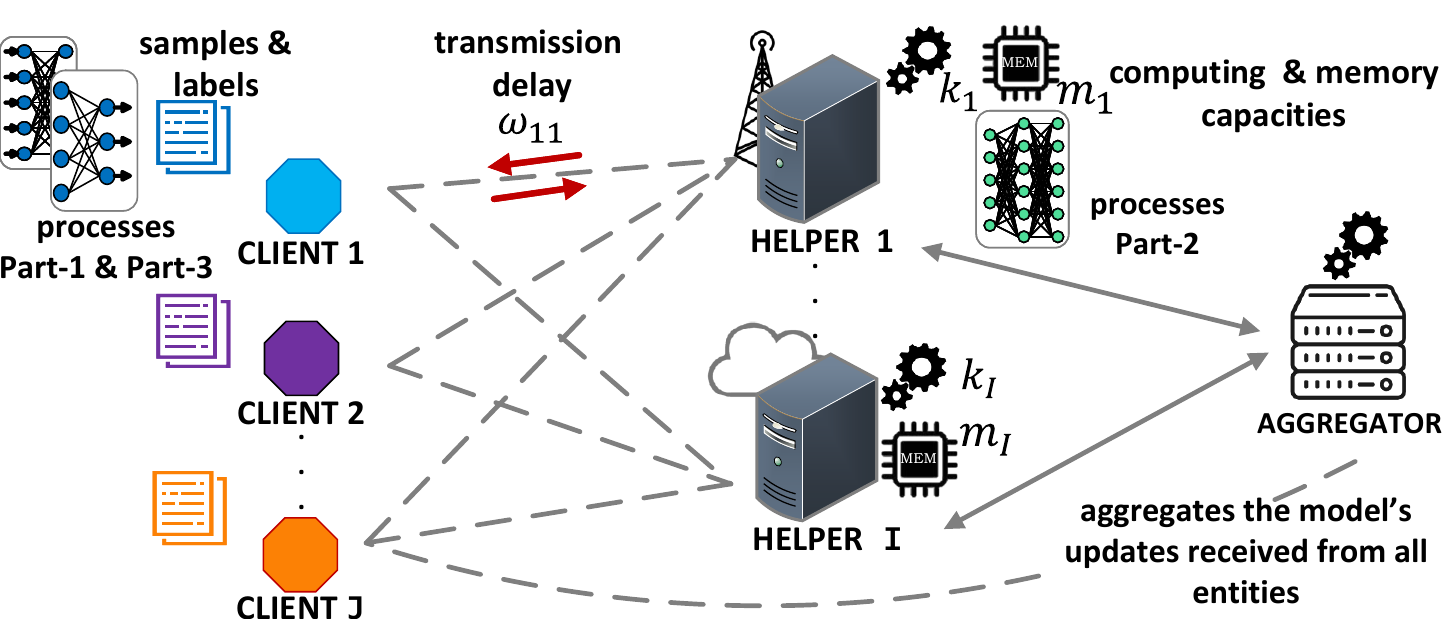} 
	\caption{Parallel SL in this work. The considered network topology, its resources, and the processing tasks per entity.\vspace{-0.3cm}}  
	\label{fig:NN+netw_topology}
\end{figure}

The presence of multiple helpers working in parallel can further reduce the training makespan~\cite{wang2023coopfl}.
Orchestrating the workflow of parallel SL  in this network of entities~(as depicted in Fig.~\ref{fig:NN+netw_topology})
is one of the main challenges that the SL paradigm faces, as discussed in~\cite{lim2020federated}. In detail, the factors that one needs to consider are the computation and memory resources of all entities,  and the connectivity of the clients to the helpers.

\textbf{Methodology \& Contributions.} Driven by the time measurements of our testbed and their heterogeneity even among devices of similar capabilities, we identify two key decisions: 1) the \emph{client-helper assignments} that are tied to the helpers' memory and computing capacities;  2) the \emph{scheduling}, i.e., the order in which each helper processes the offloaded tasks. Both decisions can be crucial for the training makespan by alleviating the effect of stragglers while fully utilizing the available resources.
Hence, we formulate the problem of \emph{jointly} making these decisions with the goal of minimizing the training makespan. To the best of our knowledge, this is the first work that studies this joint problem. We analyze this problem and its challenges both theoretically (proving it is NP-hard) and experimentally (using measurements from a realistic testbed).  Therefore, we propose a solution method based on an
intuitive decomposition of the problem into two subproblems, leveraging its inherent symmetry. The first one involves the assignment and the forward-propagation scheduling variables, and the second one involves the backward-propagation scheduling variables. 
For the former, the  Alternating Direction Method of Multipliers (ADMM) is employed, while for the latter, a polynomial-time algorithm is provided. Moreover, we propose a second solution method based on load balancing, that is more scalable, and thus, ideal for large problem instances.
Finally, our numerical evaluations provide insights on the performance of the proposed methods, as well as the achieved gains in makespan in representative scenarios. 
 The contributions of this work are summarized below.

$\bullet$ We formulate the joint problem of client-helper assignments and scheduling decisions with the goal of minimizing the training makespan in parallel SL, and prove it is NP-hard.
 
$\bullet$  We propose a solution method that is based on the decomposition of the problem into two subproblems. For the first one,  ADMM is employed, while for the second one, a polynomial-time algorithm is provided. 

$\bullet$ We extend our model to account for preemption costs, and propose a balanced-greedy algorithm with minimal overheads.

$\bullet$ We perform numerical evaluations of the proposed methods using collected data from our testbed.\footnote{The evaluation code and the collected testbed's measurements are publicly available at \url{https://github.com/jtirana98/SFL-workflow-optimization}.} These findings shape a solution strategy based on the scenario at hand. 
 
$\bullet$ We show that our solution strategy finds a near-optimal solution and achieves a shorter makespan than the baseline scheme by up to 52.3\%, and on average by $23.4\%$. 
Finally, we assess the impact of the number of helpers on the makespan. 

%% file: sections/related.tex
\section{Related Work} \label{sec:related}

\textbf{Client-based Distributed ML.} Research on FL mainly focuses  on achieving good accuracy while minimizing the wall-clock time, through client selection strategies~\cite{jiang2022adaptive}, or aggregation algorithms~\cite{rodio2023federated, chen2023gift}, or by taking into account the communication overhead~\cite{liu2023communication, jiang2022adaptive}.  
However,  some clients might not be able to support the computation demands of such protocols, which is a less explored problem. Literature on SL tackles exactly this issue~\cite{vepakomma2018split, liu2022energy, wang2021hivemind, thapa2022splitfed, tirana2022role}. A large body of existing works model a system consisting of multiple clients and a single helper. In particular, it focuses on finding the NN's cut layers while trying to optimize the energy consumption\cite{kim2023bargaining, samikwa2022ares}, or the training makespan \cite{samikwa2022ares, wu2023split}, or privacy~\cite{zhang2023privacy}. 
In presence of multiple clients, the system may need to be scaled up in order to speed up the training process. In such systems,  with multiple helpers, minimizing the training makespan requires a careful workflow orchestration.
Close to this idea, the work in \cite{wang2023coopfl} jointly finds the cut layers and assignment decisions without taking into account the scheduling decisions. As our analysis shows, scheduling decisions are crucial in systems of highly heterogeneous network resources. 
Hence, it is clear that one needs to jointly optimize the client-helper assignments and scheduling decisions in parallel SL. 

\textbf{Workflow Optimization.} Joint problems of assignment and scheduling decisions, such as the parallel machine scheduling problem, are often 
NP-hard, see, e.g.,~\cite{lawler1993sequencing, lenstra1990approximation,  chen1998review}.
While a first approach would be to rely on methods such as branch-and-bound or column or row generation methods~(like benders decomposition~\cite{geoffrion1972generalized}), our experiments show that such methods may lead to high computation overheads, even for small problem instances. Different from this approach or other existing approaches~
(e.g., for edge computing policies~\cite{10024308, 10184150}), we decompose the problem based on the inherent structure of parallel SL operations. Next, we solve one of the resulting subproblems with  ADMM   from the toolbox of convex optimization~\cite{boyd2011distributed}.  This iterative method has been recently found to perform remarkably well for non-convex problems~\cite{diamond2016general, leng2018extremely}. The advantage of employing this method lies in its versatility, allowing us to use techniques that may constrain the problem's solution space or tune its penalty parameters~\cite{diamond2016general}, and thus, we tailor it to leverage the nature of the subproblem at hand. 

%% file: sections/model.tex
\section{System Model}
\label{sec:system_model}

\textbf{Network Topology.} We consider a system with a set $\mathcal{J}$ of $J\!=\!|\cal J|$ clients, e.g., IoT or handheld 
devices, and a set $\mathcal{I}$ of $I\!=\!|\cal I|$ helpers, that are connected over a wireless bipartite network $G\!=\!(\mathcal{J}, \mathcal{I}, \mathcal{E})$ with non-interfering links $\cal E$, see Fig.~\ref{fig:NN+netw_topology}. The nodes are potentially heterogeneous in terms of hardware and/or wireless connectivity.  Namely, each node $n\in \mathcal{N}\!\!:=\!\mathcal{J}\cup \mathcal{I}$ has computing capacity $k_n$~(cycles/sec) and memory capacity $m_n$  Gbytes. 
Further, we denote $\omega_{ji}$ the average delay\footnote{In OFDMA-based systems, e.g., mobile networks, these assumptions are satisfied by design. In shared-spectrum systems, these parameters capture the effective (accounting for collisions) average delay.} for transmitting one byte from client $j$ to helper $i$, $\forall (j,i)\in \mathcal E$, and, w.l.o.g., we assume symmetric links, i.e., $\omega_{ij}\!=\!\omega_{ji}$. All nodes are connected to an aggregator, indexed  $n\!=\!0$, who may collect the necessary information and orchestrate the workflow using the solution strategy we develop.

\textbf{Parallel SL.} The clients collaborate with the helpers to train a large NN using SL and FL. Each client owns a dataset that is divided into batches of equal size.
 As discussed in Sec.~\ref{sec:intro}, the NN   is split into three parts, where $\sigma_1$ and $\sigma_2$ are the cut layers and each client computes part-1 and part-3, and offloads part-2 to a helper\footnote{An interesting approach would be to split part-2 into more parts that are offloaded and processed by more than one helper.  However, this would require a non-trivial coordination and communication among the helpers that we plan to address in future work.}. 
 This SL architecture~\cite{vepakomma2018split} protects the privacy of the clients’ data since the samples and labels are kept locally. Finally, our analysis is oblivious to: a) the cut layers, which are decided in advance and may differ across the clients, and b) the training hyperparameters~(e.g., batch size, learning rate, etc.), 
 and thus, the resulting model accuracy is not affected.

\begin{figure}[t]
	\centering  
	\includegraphics[width=9.24cm, trim={4.15cm 3.68cm 1.95cm 6.15cm},clip]{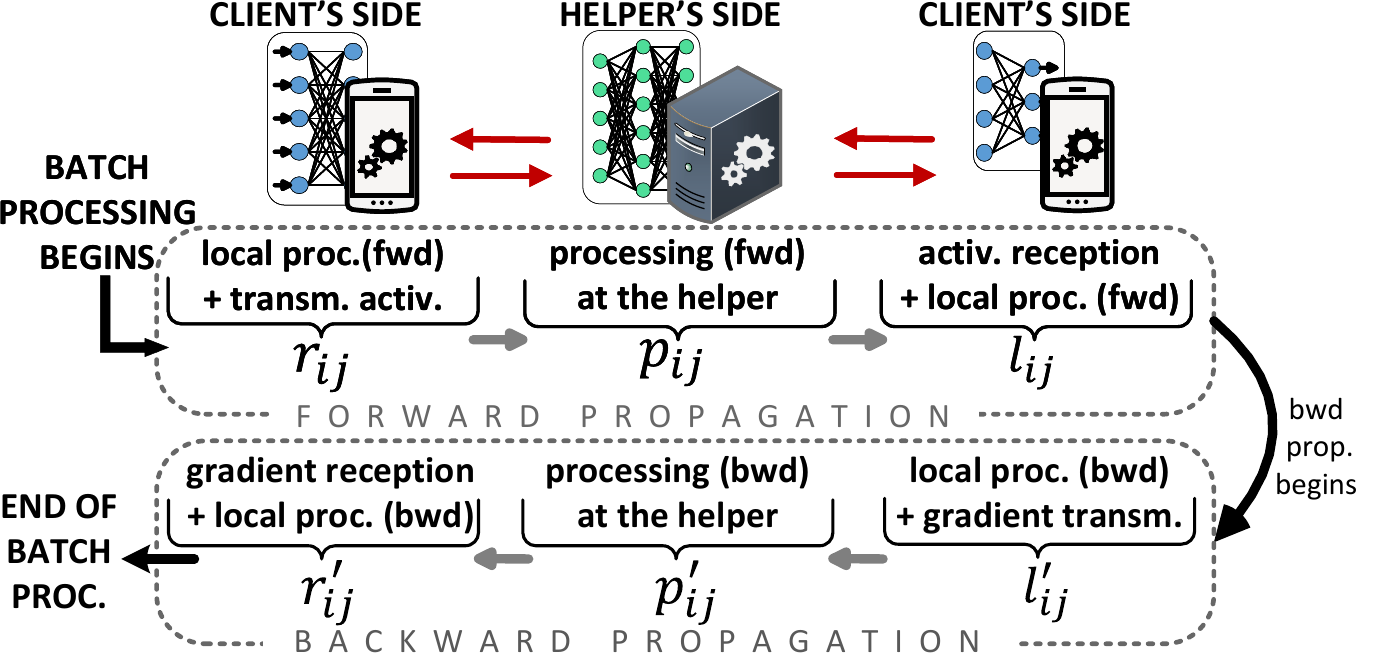} 
	\caption{The workflow of the batch processing for a single client and helper pair, 
 and the corresponding times~(processing and transmission). The \emph{queuing delay} that a client might experience at the helper is not depicted here.\vspace{-0.5cm}}
	\label{fig:times_batch_proc}
\end{figure}

\textbf{Batch Processing Workflow.} Fig.~\ref{fig:times_batch_proc}   depicts the steps of one \emph{batch update} for client $j\!\in\!\mathcal{J}$ and helper $i\!\in\mathcal I$, and introduces the main time-related parameters of parallel SL. We employ a time-slotted model~\cite{schulz2002scheduling} with time intervals that, w.l.o.g., are of unit-length.\footnote{We further discuss the choice of the time slot's length in Sec.~\ref{sec:problem} and \ref{sec:eval}.} The client applies forward-propagation of part-1 and transmits the activations of the first cut layer ($\sigma_1$) to the helper. We denote by $r_{ij}$ the number of time slots required for these two operations, which depends on $k_j$ and $\omega_{ij}$. The helper needs $p_{ij}$ time slots to propagate these activations into part-2~(i.e., to execute the \texttt{fwd-prop} task), which depend both on the capacity $k_i$ and the choice of cut layers,~i.e., the ``size'' of the task. The client receives the activations of the last layer of part-2 ($\sigma_2$) from the helper and completes forward-propagation by processing part-3 and computing the loss. We denote by $l_{ij}$ the time required for these operations, which depends on $\omega_{ij}$, the data size, and $k_j$. 

Then, the back-propagation of the training error starts. The client updates the weights of part-3, computes the gradients, and transmits them to the helper, consuming $l_{ij}^\prime$ time slots. The helper back-propagates these gradients into part-2, so as to update its weights, spending $p_{ij}^\prime$ time to execute this \texttt{bwd-prop} task. Afterwards, it transmits the gradients  of $\sigma_1$  to the client, who then back-propagates part-1. We denote by $r_{ij}^\prime$ the time required for this final step.  The time-related parameters (or delays)   $\bm r\!=\!(r_{ij}, (i,j)\in \cal E)$, $\bm r^\prime\!=\!(r_{ij}^\prime, (i,j)\in \cal E)$, $\bm p\!=\!(p_{ij}, (i,j)\in \cal E)$, $\bm p^\prime\!=\!(p_{ij}^\prime, (i,j)\in \cal E)$, $\bm l\!=\!(l_{ij}, (i,j)\!\in\! \cal E)$, and $\bm l^\prime\!=\!(l_{ij}^\prime, (i,j)\!\in\! \cal E)$ represent average quantities\footnote{As is common in scheduling literature and, w.l.o.g., we assume that these  quantities are integers. If this assumption is violated, fractions can be handled by multiplying by a proper factor. Also, one could adopt a more conservative approach where worst-case values are considered instead of the average ones.}
for  these tasks or processes,   and are considered available through profiling and other offline measurement methods~\cite{tran2019federated, fu2022kalmia}.

\textbf{Epochs \& Aggregation.} The batch processing workflow  is repeated for all batches. When the client has applied a batch update using all batches of data, a \emph{local epoch} is completed. Clients repeat the processing of a local epoch for a predefined number of times until a \emph{training round}~(or global epoch) is completed. Subsequently, the updated model parts from each node (client or helper) must be sent to and  aggregated at the aggregator, using methods such as $\mathtt{FedAvg}$~\cite{mcmahan2017communication}. 
Typically, such training processes require hundreds of training rounds, each consisting of multiple batch updates \cite{samikwa2022ares,thapa2022splitfed}. Hence, in order to minimize the maximum training time across all clients, i.e., the  \emph{training makespan}, we leverage the structural nature of the training process,  and focus on  the makespan of a single batch,  see~\cite{samikwa2022ares,wang2023coopfl}. We note that, when compared to  conventional FL, the time required for the aggregation~(comprising processing and transmissions) is negligible
since the size of the data transferred is smaller per entity. Moreover, transmitting model updates can start even before all entities have completed a batch update, thus speeding up the procedure. 

It is important to stress the  inherent coupling between  forward and backward propagation. When a client $j\in \cal J$ transmits part-1 activations to a helper $i\in\cal I$, the latter allocates $d_j$  Gbytes 
of memory, where possibly $d_j\neq d_{j^\prime}$ if $j\neq j^\prime$, in order to store and process these activations. The helper stores this data during the \texttt{fwd-prop}, and reassigns the (same) memory to the gradients received from the client during the \texttt{bwd-prop}. This means that, in practice, a client cannot use a different helper for each propagation direction.

\textbf{Time Horizon \& Decision Variables}. As mentioned earlier, we employ a time-slotted model with time intervals  $S_t$, where $S_0\!=\![0,1]$,  $S_t\!=\!(t, t+1], t=1, \ldots, T$, and $T\!=\!|\mathcal T|$ is the number of slots in time horizon $\mathcal T$. The parameter $T$ upper-bounds the batch makespan, and can be calculated as follows:
\begin{equation}
    T:= \max_{(i,j)\in\mathcal E} \left\{ r_{ij} + l_{ij} + r_{ij}^{\prime} +  l_{ij}^{\prime} \right\} + \sum_{j\in\cal J} \max_{i\in\mathcal I} \left\{p_{ij} + p_{ij}^{\prime} \right\}, \notag
\end{equation}
where the first term finds the worst-case transmission, and processing times in the network; and the second term measures the worst helper's processing time for any task. 

Based on this time-slotted model, we introduce variables that 
inject tasks to helpers towards minimizing the makespan. In particular, we introduce the binary variables $\bm y=(y_{ij}\in \{0,1\}, (i,j)\in\cal E)$, where $y_{ij}=1$ if client $j$ is assigned to  helper $i$. Moreover, we define the slot-indexed variables $\bm x=(x_{ijt} \in \{0,1\}, (i,j)\in\mathcal{E}, t\!\in\!\mathcal{T})$, where $x_{ijt} =1$ if the \texttt{fwd-prop} task of client $j$ is  processed at helper $i$ during slot $S_t$, and $x_{ijt} =0$ otherwise. Similarly, we define $\bm z=(z_{ijt}\in \{0,1\}, (i,j)\in\mathcal{E}, t\!\in\!\mathcal T)$ with $z_{ijt}=1$ if the  \texttt{bwd-prop} task of client $j$  is processed  at $i$ during $S_t$. These vectors fully characterize the batch processing workflow.

%% file: sections/problem.tex
\section{Problem Formulation} \label{sec:problem}

The scheduling and assignment decision variables  ($\bm x, \bm z$, and $\bm y$) should be consistent with the   SL operation principles.

\textbf{Scheduling Constraints}. Each \texttt{fwd-prop} task can be executed only after its input becomes available, i.e., after activations of $\sigma_1$ are transmitted to the helper. Hence,
\begin{equation}
 \label{eq:constraint_x_def}
 x_{ijt} = 0, \quad \forall t<r_{ij}, \ (i, j)\in \mathcal{E}.
\end{equation}
In scheduling parlance, $r_{ij}$ is the \emph{release time} of this task, i.e., when it becomes ``available'' at the helper. Similarly,  the \texttt{bwd-prop} can start only after the gradients of $\sigma_2+1$ have been received by the helper~(see Fig.~\ref{fig:times_batch_proc}), and thus,
\begin{equation}\label{eq:preced-2}
z_{ij(t+l_{ij}+l_{ij}^{\prime})} \leq \frac{1}{p_{ij}} \sum_{\tau =0}^{t-1} x_{ij\tau}, \quad \forall (i,j)\in \mathcal{E}, t\in \mathcal T.
\end{equation}
That is, in order to assign the \texttt{bwd-prop} task of $j$ to $i$ at (or after) slot $t+l_{ij}+l_{ij}^{\prime}$, we need to allocate enough processing time at $i$ (at least $p_{ij}$) for \texttt{fwd-prop} until slot $t$. Essentially, \eqref{eq:preced-2} are  \emph{precedence constraints} that ensure the \texttt{bwd-prop} of a client's part-2 strictly succeeds its \texttt{fwd-prop}. The next constraints ensure that each helper will process a single task during any time slot (assuming single-threaded computing):
\begin{equation} \label{eq:one_at_time}
\sum_{j\in\cal J} \left( x_{ijt} + z_{ijt} \right) \leq 1, \quad \forall i\in \mathcal{I}, t\in \mathcal T.
\vspace{-0.1cm}
\end{equation}

\textbf{Assignment Constraints}. Regarding the assignment decisions $\bm y$, each client's task is assigned to a single helper: 
\begin{equation} \label{eq:single_machine}
\sum_{i\in\mathcal I} y_{ij}=1, \quad  \forall j\in \mathcal{J}. \vspace{-0.1cm}
\end{equation}
Further, the assignments are bounded by the helper's memory:
\begin{equation} \label{eq:constraint_memory}
\sum_{j\in \mathcal J} y_{ij}d_j \leq m_i, \quad  \forall i\in \mathcal{I}, \vspace{-0.1cm}
\end{equation}
and recall that this memory is used in both directions. 
Clearly, 
the assignment and scheduling constraints are tightly coupled. Indeed, when an assignment is decided, we need to ensure adequate processing time will be scheduled for the  \texttt{fwd-prop} and \texttt{bwd-prop} tasks. In other words, it should hold that:
\begin{align}
&\sum_{t\in\mathcal T} x_{ijt} \! =\! y_{ij} p_{ij}, \quad \forall (i,j)\in\mathcal{E} \ \  \text{and} \label{eq:newcontraint_xp} \\     
&\sum_{t\in\mathcal T} z_{ijt} \!=\!  y_{ij} p_{ij}^{\prime} , \quad \forall (i,j)\in\mathcal{E}. \label{eq:newcontraint_yp}
\end{align}

\textbf{Completion Times}. Finally, we introduce additional variables to measure some key delays. In particular, we define $\bm \phi=(\phi_j, j\in\mathcal J)$, where $\phi_j$ is the slot when the \texttt{bwd-prop} of client $j\in \mathcal J$ is completed. 
These variables should satisfy:
\begin{equation}\label{eq:fbwd}
    \phi_j \geq (t+1) z_{ijt}, \quad \forall (i,j)\in\mathcal{E}, t\in\mathcal{T}.
\end{equation}
Similarly, we introduce the overall (batch) completion time variable $c_j, \forall j\in \mathcal J$, which should satisfy:
\begin{equation}\label{eq:completition}
    c_j= \phi_j + \sum_{i\in\mathcal{I}} r_{ij}^{\prime} y_{ij}  \quad \forall   j\in \mathcal{J}.
\end{equation}
Essentially, the vector $\bm c\!=\!(c_j, j\in\mathcal J)$ contains the completion times of one-batch model-training for all clients, and hence, its maximum element dictates the makespan. Naturally, all elements of $\bm \phi$ and $\bm c$ are upper-bounded by $T$. Finally, we observe that the quantity $\phi_j-\sum_i y_{ij}(r_{ij}+p_{ij}+l_{ij}+l_{ij}^\prime +p_{ij}^\prime)$ is the total queuing delay that client $j$ might experience during \texttt{fwd-prop} and \texttt{bwd-prop}.

\textbf{Preemption.} A strong aspect of our model is that it allows preemption, i.e.,  a task may be paused partway through its execution and then resumed later, if this improves the makespan. Specifically, preemptions may occur at the end of each time slot $S_t$.  Preemptive schedules may prioritize 
the slowest client~(straggler), thus reducing the makespan. 
This is in contrast to previous work that follows more rigid non-preemptive models~\cite{wang2023coopfl}, but in line with related work on edge computing, e.g.,  \cite{fu2022kalmia}, \cite{meng2019dedas}. We further discuss this point in Sec.~\ref{sec:discussion_extentions}.
Finally, since the length of $S_t$ determines the frequency of preemptions, a smaller length implies a larger benefit from preemption, i.e., shorter makespan. We investigate this point using our testbed's measurements in Sec.~\ref{sec:eval}.

We can now formulate the \emph{joint scheduling and assignment} problem that minimizes the batch makespan of parallel SL. 
\theoremstyle{definition} \newtheorem{problem_fwd_bwd}{Problem}
\begin{problem_fwd_bwd}[Batch Training Makespan]
	\begin{IEEEeqnarray}{rclll} 
			\mathbb{P}: 	  &\underset{\boldsymbol{x,z,y, \phi, c}}{\text{ minimize  }}&  \max_{j\in\mathcal J}\left\{ c_j \right\} &&\nonumber \IEEEeqnarraynumspace \\ 
		&\text{s.t.  }& (1)-(9),\notag \\
		&& \bm x, \bm z \in \{0,1\}^{|\cal E|\times |\cal T| }, \bm \phi, \bm c\in \{0 \twodots T \}^{J},    \IEEEeqnarraynumspace  \\
  && \bm y\in \{0,1\}^{|\cal E |}.  \label{eq:prob1_y_binary}\IEEEeqnarraynumspace  
	\end{IEEEeqnarray}
\end{problem_fwd_bwd}
\noindent This min-max problem can be written as an Integer Linear Program~(ILP) using standard transformations~\cite[Sec. 4.3.1]{boyd2004convex}, i.e., by introducing the worst-case makespan variable $\xi$ and changing the objective to $\min_{ \xi, \bm x,\bm y,\bm z, \bm \phi} \xi $ with additional constraints $\xi \!\geq \!c_j, \forall j\!\in\! \mathcal J$. Albeit elegant, such transformations do not alleviate the computational challenges in solving large, or even medium-sized instances of  $\mathbb P$. To exemplify, for a scenario with $J\!=\!20$ clients, $I\!=\!5$ helpers, and horizon $T\!=\!636$, state-of-the-art  solvers, such as Gurobi~\cite{gurobi}, achieve only a $40\%$ optimality gap in $14$ hours (off-the-shelf computer). Such long delays are not surprising due to the following result. 
\theoremstyle{plain} \newtheorem{lemma:hard}{Theorem} 
\begin{lemma:hard} \label{lemma:hard}
$\mathbb{P}$~(Problem 1) is NP-hard.
\end{lemma:hard}
\begin{proof}
We first define a simpler  instance of $\mathbb{P}$: we assume that all helpers have enough memory for all tasks, i.e., we drop the memory constraints in \eqref{eq:constraint_memory}, and $\bm r = \bm r^\prime =\bm l = \bm l^\prime = \bm 0$, i.e., transmissions and propagations of part-1 and 3 are instantaneous. Therefore, all \texttt{fwd-prop} tasks are released at time 0 and each \texttt{bwd-prop} task may start instantly  after the \texttt{fwd-prop} task is processed. Moreover, we assume that $\bm p = \bm p^\prime = \bm 1$, i.e., all tasks are of unit-length (require one time slot). We will show that there is a polynomial-time reduction from the parallel machine scheduling problem in~\cite{ullman1967complexity} to this problem. The former is defined as follows: given a set of $n$  jobs and a set of $m$   parallel machines, each job should be assigned to a machine, while every machine can process at most one job at a time. The processing time of all jobs is  $q=1$, jobs are subject to precedence constraints,  and the problem has as objective to find a schedule and the job-machine assignments to minimize the makespan, i.e., the time the last job will be completed at the machine. The reduction is shown by setting $n=J$, i.e., the \texttt{fwd-prop} and \texttt{bwd-prop} are the jobs (with precedence constraints), $m=I$, i.e., the helpers are the machines, and $p_{ij}=p_{ij}^\prime =1, \; \forall (i,j)\in\mathcal{E}$. Given this reduction and the fact that the parallel machine scheduling problem is NP-hard~\cite{ullman1967complexity, lawler1993sequencing}, $\mathbb{P}$ is NP-hard as well.   \end{proof}

Given this result, we develop a multi-fold solution strategy consisting of a decomposition algorithm~(Sec.~\ref{sec:solution}) and an informed heuristic~(Sec.~\ref{sec:discussion_extentions}).

\section{Solution method}
\label{sec:solution}
\begin{figure}[t]
	\centering  
\includegraphics[width=7.5cm, trim={0.4cm 9.52cm 2.38cm 2.75cm},clip]{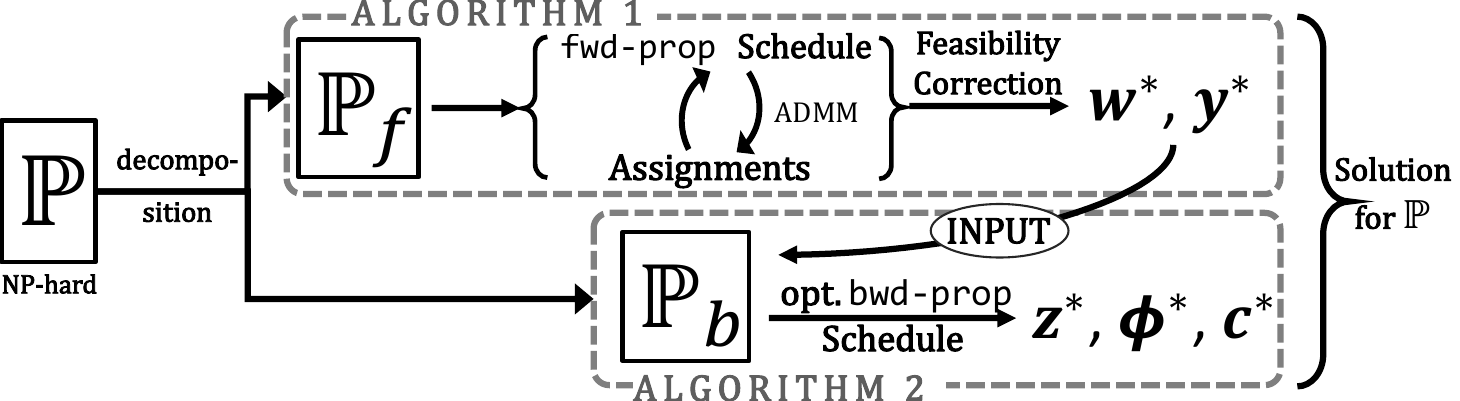} 
	\caption{The roadmap to our ADMM-based solution method. \vspace{-0.3cm}}  
	\label{fig:roadmap}
\end{figure}

The core idea of our solution method is to decompose $\mathbb P$ into two subproblems (see Fig.~\ref{fig:roadmap}): \emph{(i)} $\mathbb P_f$, which minimizes the forward propagation makespan by deciding variables $\bm x$ and $\bm y$; \emph{(ii)}  $\mathbb P_b$, which minimizes the backward makespan by deciding $\bm z, \bm \phi, \bm c $.  We solve $\mathbb P_f$ using ADMM (see discussion in Sec.~\ref{sec:related}), and, for $\mathbb P_b$,  we prove it admits a polynomial-time algorithm by leveraging its coupling with $\mathbb P_f$ (due to $\bm y$). As we will see in Sec.~\ref{sec:eval}, this approach will lead to considerable speedups~(up to $52\times$) when compared to exact solution methods.

\subsection{\texttt{Fwd-prop} Optimization}

Before introducing $\mathbb P_f$, we need some additional notation. First, we note that the time horizon that is related to \texttt{fwd-prop} can be confined to the set $\mathcal T_f$ with $T_f:= \max_{(i,j)\in\mathcal E} \{ r_{ij} + l_{ij}\}  + \sum_{j\in\mathcal J} \max_{i\in\mathcal I} p_{ij}$. We denote by $\phi_j^f$ the \texttt{fwd-prop} finish time for each client $j\in\mathcal J$, which by definition has to satisfy the constraints~(similarly to~\eqref{eq:fbwd}):
\begin{align}\label{eq:fwd_finish}
\phi_j^{f} \geq (t+1) x_{ijt}, \ \forall i\in\mathcal I, t\in\mathcal{T}_f.
\end{align}
Also, we define the \texttt{fwd-prop} completion time $c_j^f, j\!\in\!\mathcal J$, which is determined by   $\phi_j^{f}$ and the times $l_{ij}$, i.e., 
\begin{align}\label{eq:fwd_completion}
c^{f}_j = \phi_j^{f}  + \sum_{i\in\mathcal I} l_{ij}y_{ij},  \ \forall j\in\mathcal J.
\end{align}
As before, $\bm \phi^f=(\phi_j^{f}, j\in\mathcal J)$ and 
$\bm c^f\!=\!(c_j^f, j\in\mathcal J)$. 
Collecting the above requirements, we can now formulate $\mathbb P_f$:
\theoremstyle{definition} \newtheorem{problem_fwd_only}[problem_fwd_bwd]{Problem}
\begin{problem_fwd_only}[\texttt{Fwd-prop} makespan]
	\begin{IEEEeqnarray}{rclll} 
	\mathbb{P}_f: &\underset{\boldsymbol{x,y,\phi^f,c^f}}{\text{ minimize  }}&  \max_{j\in\mathcal J}\big\{ c^{f}_j\big\} &&\nonumber \IEEEeqnarraynumspace \\ 
		&\text{s.t.  }& \eqref{eq:constraint_x_def}, \eqref{eq:single_machine}-\eqref{eq:newcontraint_xp}, \eqref{eq:prob1_y_binary}, \eqref{eq:fwd_finish}, \eqref{eq:fwd_completion} \nonumber &&\\
		&& \sum_{j\in\mathcal J}  x_{ijt}  \leq 1, \quad \forall i\in\mathcal I, t\in\mathcal{T}_f,& \label{eq:fwd_one_job_per_interval} \\	
		&& \bm x \in \{0,1\}^{|\cal E|\times |\cal T| }, \bm \phi^f, \bm c^f\in \{0\twodots T_f\}^{J}. &&\IEEEeqnarraynumspace  \label{eq:binary-var-fwd-constraint-a}		 
	\end{IEEEeqnarray}
\end{problem_fwd_only}
Comparing the constraints of this reduced problem with $\mathbb P$, we  observe that: $\mathbb P_f$ replaces constraints \eqref{eq:fbwd} and \eqref{eq:completition} with \eqref{eq:fwd_finish} and \eqref{eq:fwd_completion};  omits constraints \eqref{eq:preced-2} and \eqref{eq:newcontraint_yp}; and replaces \eqref{eq:one_at_time} with \eqref{eq:fwd_one_job_per_interval}. This yields a simpler problem, as $\mathbb{P}_f$ has fewer variables (omits $\bm z$ and $T_f<T$) 
and less complicated constraints. However, the solution of $\mathbb P_f$ will not necessarily be consistent with the \texttt{bwd-prop} operations. For this, we properly tune the \texttt{bwd-prop} scheduling problem $\mathbb P_b$  in Sec. \ref{sec:bwd-prop-optimization}. Despite this decomposition,   there is not a known algorithm for $\mathbb P_f$. In fact,  arguments as the ones in the proof of Theorem~\ref{lemma:hard}  can lead to a reduction from the unrelated machine scheduling problem with release dates, preemption, and no precedence constraints to $\mathbb{P}_f$. To the best of our knowledge, there is no polynomial-time algorithm for this problem, except for some special cases, e.g., for a specific number of machines, see~\cite{lawler1993sequencing, gonzalez1990optimal, horn1974some}. 

To that end, we employ ADMM to decompose $\mathbb P_f$ and obtain smaller subproblems, which, it turns out, can be solved in reasonable time for many problem instances. Indeed, we observe that by relaxing the constraints in \eqref{eq:newcontraint_xp}, we can decompose $\mathbb P_f$ into a (forward-only) scheduling subproblem, involving $(\bm x, \bm \phi^f, \bm c^f)$, and an assignment subproblem that optimizes $\bm y$. Then, we can solve these subproblems iteratively and penalize their solution so as to gradually recover the relaxed constraints. The first step is to define the Augmented Lagrange function:
\begin{IEEEeqnarray}{lll}
   \mathcal L(\boldsymbol{w}, \bm y, \boldsymbol{\lambda}) =&&  \max_{j\in\mathcal J} c^{f}_j + \sum_{(i,j)\in\mathcal E} \lambda_{ij} \bigg( \sum_{t\in\mathcal T_f} x_{ijt} - y_{ij}p_{ij} \bigg) \nonumber \IEEEeqnarraynumspace \\ &&+ \frac{\rho}{2} \sum_{(i,j)\in\mathcal E} \Big|\sum_{t\in\mathcal T_f} x_{ijt} - y_{ij}p_{ij} \Big|, \IEEEeqnarraynumspace
\end{IEEEeqnarray}
where we define $\bm w\!=\!(\bm x, \bm \phi^f, \bm c^f)$ to streamline the notation; introduce the dual variables $\bm \lambda=(\lambda_{ij}, (i,j)\in\mathcal E )$ for relaxing \eqref{eq:newcontraint_xp}; and use the ADMM penalty parameter $\rho$, see \cite[Ch. 3]{boyd2011distributed}. Note that, unlike the vanilla version of ADMM that uses the Euclidean norm $\ell_2$, we penalize the constraint violation using the $\ell_1$ norm so as to improve the algorithm runtime. 

\begin{algorithm}[t]
\NoCaptionOfAlgo

\SetAlgoLined
\SetKwInOut{Input}{Input}
\SetKw{Return}{Return}
\SetAlgoLined

\Input{$\bm \lambda^{(0)},  \bm y^{(0)} =\bm 0$, $\varepsilon_1$, $\varepsilon_2$, $\rho$, $\tau_{max}$, $T^f$}
 \For{$\tau=1,2,\ldots, \tau_{max}$}{$\bm w^{(\tau+1)}\!=\! \underset{\footnotesize{\eqref{eq:constraint_x_def}, \eqref{eq:fwd_finish}-\eqref{eq:binary-var-fwd-constraint-a}, \eqref{eq:schedule-time-new}}}\argmin\!\mathcal{L}\big(\bm w, \! \bm y^{(\tau)}\!, \bm{\lambda}^{(\tau)}\big)$\! \footnotesize{\emph{schedule}}\\
 \normalsize
 $\bm{y}^{(\tau+1)}=\!\underset{\eqref{eq:single_machine},\eqref{eq:constraint_memory}, \eqref{eq:prob1_y_binary} }\argmin \mathcal{L}\big(\bm{w}^{(\tau+1)}, \bm{y}, \bm{\lambda}^{(\tau)}\big)$ \footnotesize{\emph{assignment}}\\
$\lambda_{ij}^{(\tau+1)} =  \lambda_{ij}^{(\tau)}+ \Big(\sum_{t\in\mathcal T_f}  x_{ijt}^{(\tau+1)} - 	y_{ij}^{(\tau+1)}p_{ij}\Big),  \ \ \forall (i,j)\in \mathcal{E}$\\
\normalsize
Exit \textbf{for} if  \eqref{eq:converg-1} and \eqref{eq:converg-2} are satisfied.} Correct $\mathbb P_f$ feasibility with \eqref{eq:fwd-feasib-correction}.
 
 \Return{$\bm w^*\!=\!(\bm x^*, \bm \phi^{f*}, \bm c^{f*})$, $ \bm y^*$}
 \caption{\textbf{Algorithm 1} ADMM-based \texttt{fwd-prop} Workflow} \label{alg:fwd}
 \vspace{-0.1cm}
\end{algorithm}

The detailed steps can be found in Algorithm \ref{alg:fwd}. At iteration $\tau +1$,   we first update the schedule $\bm w$ using the previous assignment $\bm y^{(\tau)}$ and dual variables $\bm \lambda^{(\tau)}$~(line 2). Next, we optimize the assignment $\bm y^{(\tau + 1)}$ using the updated schedule $\bm w^{(\tau+1)}$~(line 3), and finally we correct the dual variables based on the violation of~\eqref{eq:newcontraint_xp} in line 4. We repeat these steps until convergence is achieved or a maximum number of iterations is reached~($\tau_{max}$).\footnote{The $\bm w$- and $\bm y$-subproblems (i.e., lines 2-3 of Alg.~\ref{alg:fwd}) could be solved either with exact methods, e.g., branch and bound, or inexact methods, e.g., through a tailored relaxation~\cite{leng2018extremely}. As for the former, we elaborate on the resulting overhead in Sec.~\ref{sec:eval}, and  the latter is in line with the fact that ADMM  can (under certain
conditions)  tolerate inexact solutions for its subproblems~\cite{eckstein1992douglas}.\label{foot_1}} As a convergence flag, we use the detection of stationary assignments and objective values~(line 5): 
\begin{align}
&\sum_{(i,j) \in\mathcal E}\!\Big|y_{ij}^{(\tau+1)}\!-\!y_{ij}^{(\tau)}\Big|\!<\!\varepsilon_1 \  \  \text{and} \label{eq:converg-1} \\
&\ \Big|\max_{j\in\mathcal J}c_j^{f, (\tau+1)}\!-\!\max_{j\in\mathcal J}c_j^{f, (\tau)}\Big|\!<\!\varepsilon_2. \label{eq:converg-2}
\end{align}

\noindent Finally, we correct any remaining infeasible constraints by tuning the schedule to the final assignment $\bm y^*$~(line 6):
\begin{align}\label{eq:fwd-feasib-correction}
\bm w^{*}=\underset{\eqref{eq:constraint_x_def}, \eqref{eq:fwd_finish}-\eqref{eq:binary-var-fwd-constraint-a}, \eqref{eq:newcontraint_xp}}\argmin \mathcal{L}\big(\bm w, \bm y^*, \bm{\lambda}^*\big),
\end{align}
where   we additionally use  the  constraints \eqref{eq:newcontraint_xp} to ensure full consistency between $\bm x^*$ and $\bm y^*$. Since the relaxed constraints in \eqref{eq:newcontraint_xp} concern the processing times for each client's task, we can further accelerate the convergence of Alg.~\ref{alg:fwd} by creating a tighter constraint set~\cite{diamond2016general}. In detail, we  introduce a set of constraints  that limit  the search to schedules that allocate enough processing time for each client in the $\bm w$-subproblem: 
\begin{equation}\label{eq:schedule-time-new}
\vspace{-0.1cm}
	\sum_{i\in\mathcal I} \frac{1}{p_{ij}}\sum_{t\in\mathcal T_f} x_{ijt} =1, \quad \forall j \in \mathcal{J}.
\end{equation}

Algorithm \ref{alg:fwd} is not guaranteed to converge to the optimal solution of $\mathbb P_f$. However, its efficacy is demonstrated with a battery of trace-driven evaluations in Sec.~\ref{sec:eval}, where
in most of the tested scenarios, it achieves less than $10.2\%$ suboptimality gap, with one corner case of $14.9\%$.

\subsection{\texttt{Bwd-prop} Schedule} \label{sec:bwd-prop-optimization}

Given the assignment $\bm y^*$ and \texttt{fwd-prop} schedule $\bm w^*\!=\!(\bm x^*, \bm \phi^{f*}, \bm c^{f*})$ from the solution of $\mathbb P_f$, we can optimize the \texttt{bwd-prop} schedule by solving the $\mathbb P_b$ subproblem. The latter stems from $\mathbb P$ by removing the constraints which do not involve $\bm z$; and by further replacing variables $\bm x$ and $\bm y$ with the respective values $\bm x^*$ and $\bm y^*$ that we obtained from $\mathbb P_f$, wherever they appear in the constraints:

\theoremstyle{definition} \newtheorem{problem_bwd_decoupled}[problem_fwd_bwd]{Problem}
\begin{problem_bwd_decoupled}[\texttt{Bwd-prop} makespan; given $\bm y^*$, $\bm w^*$]
\begin{IEEEeqnarray}{rclll} 
\mathbb{P}_b: \; &\underset{\bm z, \bm \phi, \bm c}{\text{ minimize  }}&  \max_{j\in\mathcal J} c_j &&\nonumber \IEEEeqnarraynumspace \\ 
&\text{s.t.  }&  \eqref{eq:preced-2}, \eqref{eq:one_at_time},  \eqref{eq:newcontraint_yp}-
\eqref{eq:completition} && \nonumber \\
&&  \bm{z} \in \{0,1\}^{|\mathcal{E}|\times|\mathcal{T}|},  \bm{\phi}, \bm{c} \in \{T_f^*\twodots T\}^{J}. \IEEEeqnarraynumspace  
\vspace{-0.1cm}
\end{IEEEeqnarray}
\end{problem_bwd_decoupled}
\noindent We stress that the variables $\bm \phi$ and $\bm c$ can be restricted in the time window starting after the \texttt{fwd-prop}, which is provided by $\mathbb P_f$ and denoted by $T_f^*$. These provisions ensure that the solution we obtain from successively solving  subproblems $\mathbb P_f$ and $\mathbb P_b$ 
will not induce constraint violations. 

\theoremstyle{plain} \newtheorem{lemma:poly-solvable}[lemma:hard]{Theorem} 
\begin{lemma:poly-solvable}\label{lemma:poly-solvable}
$\mathbb{P}_b$  can be solved in polynomial time.
\end{lemma:poly-solvable}

\begin{figure}[t]
	\centering  
\includegraphics[height=3.5cm, trim={1.46cm 5.71cm 12.38cm 7.05cm},clip]{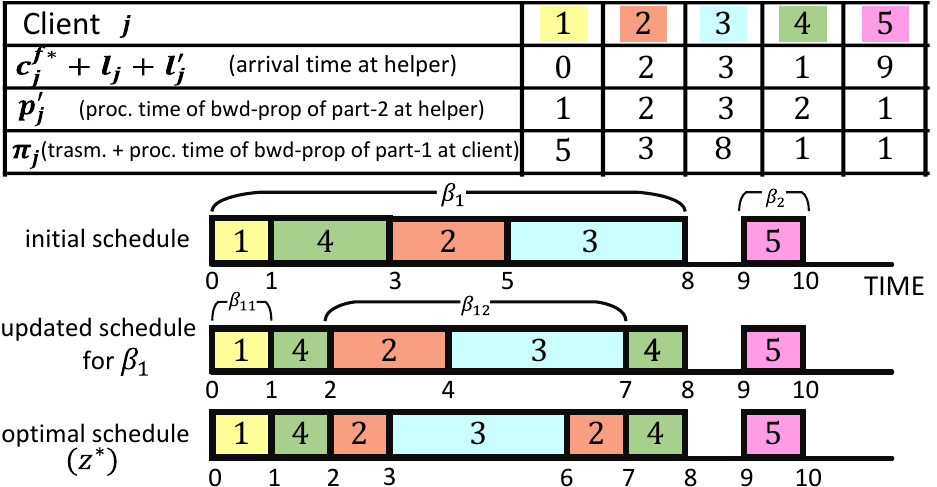} 
	\caption{Algorithm 2 for optimal \texttt{bwd-prop} schedule in a toy example of 5 clients and 1 helper. \vspace{-0.2cm}}  
	\label{fig:algo2}
\end{figure}

\begin{proof}
We first observe that, since the client-helper assignments are fixed~($\bm y^*$), we can parallelize $\mathbb{P}_b$'s solution across the helpers. That is, we can independently focus on  the \texttt{bwd-prop} tasks of the subset of clients $\mathcal J_i$ assigned to each helper $i\in\mathcal I$, where $\mathcal{J}_i:= \{j\in \mathcal{J} \; :\; y_{ij}^*=1\}$. Also,  the $\bm w^*$ obtained by Alg.~\ref{alg:fwd}   dictates a subset of time slots where \texttt{bwd-prop} tasks can be scheduled. We denote by $\mathcal{T}_{i}$ the remaining eligible slots for helper $i$. 
We can now state the subproblem of minimizing the \texttt{bwd-prop} makespan for each helper $i\in \mathcal{I}$, while we abuse notation and drop the index $i$.
\begin{IEEEeqnarray}{rclll} 
\mathbb P_b^i: &\underset{\bm z, \bm \phi}{\text{ minimize }}& \max_{j\in\mathcal J_i} \left\{ \phi_j + \pi_{j} \right\} &&\nonumber \IEEEeqnarraynumspace \\ 
&\text{s.t.}& \sum_{t\in\mathcal{T}_{j}}z_{jt}=p_{j}^\prime, \quad \forall j\in \mathcal{J}_i && \vspace{-0.1cm} \\
&&  z_{j(t+l_{j}+l_{j}^{\prime})} \leq \frac{1}{p_{j}} \sum_{\tau =0}^{t-1} x_{j\tau}^*, \; \forall j\in \mathcal{J}_i, t\in \mathcal T_i &&\IEEEeqnarraynumspace  \\
&& \sum_{j\in\mathcal J_i}z_{jt} \leq 1, \quad \forall t\in \mathcal T && \IEEEeqnarraynumspace  \\
&& \phi_j\geq (t+1)z_{jt}, \quad \forall j\in \mathcal J_i, t\in\mathcal T_i,&& \IEEEeqnarraynumspace 
\end{IEEEeqnarray}
\noindent where $\pi_{j}:=\sum_{i\in\mathcal I}r_{ij}^{\prime}y_{ij}^*, \forall j\in \mathcal{J}_i$ and $x_{j\tau}^*$ are fixed parameters. 
We will show that there is a polynomial-time reduction from $\mathbb{P}_b^i$ to the single machine scheduling problem of minimizing the maximum cost subject to release and precedence constraints, which is polynomially time solvable~\cite{baker1983preemptive}. It suffices to set the release times of the jobs as the $\{c_j^{f*}+l_j+l_j^\prime\}_{j\in \mathcal{J}_i}$ (i.e., the time the client needs to complete the back-propagation of part-3 and transmit the gradients, given the $c_j^{f*}$). Moreover, it suffices to set as the cost function in~\cite{baker1983preemptive}  the quantity $\phi_j + \pi_j$, i.e., the makespan of the batch update (including the clients' local computations).
\end{proof}
We now present the algorithm that optimally solves $\mathbb P_b$ based on~\cite{baker1983preemptive} together with a worked example in a scenario of 5 clients and 1 helper, as depicted in Fig.~\ref{fig:algo2}.

\textbf{Algorithm 2 and Worked Example.} For each helper $i$, and  \emph{in parallel}, we find the set of assigned clients $\mathcal{J}_i$. We then order the clients according to nondecreasing $\{c_j^{f*}+l_j+l_j^\prime\}_{j\in \mathcal{J}_i}$, which are the arrival~(release) times for their \texttt{bwd-prop} tasks at the helper, and build an initial schedule where tasks are processed according to the ordering of their arrival times. This schedule naturally decomposes $\mathcal{J}_i$ into an initial set $\mathcal{B}_i$ of \emph{blocks}. Specifically, each block $\beta \in \mathcal{B}_i$ is the smallest set of clients whose \texttt{bwd-prop} task has an arrival time   at (or after) $s(\beta) := \min_{j\in\beta} \{c_j^{f*}+l_j+l_j^\prime\}$ and can be processed before $e(\beta):=s(\beta) + \sum_{j\in\beta} p_j^\prime$.  In fact, a client's task $h\not\in \beta$ is either processed no later than $s(\beta)$, i.e., $\phi_h \leq s(\beta)$, or not released at the helper before $e(\beta)$, i.e., $c_h^{f*}+l_h+l_h^\prime \geq e(\beta)$. Essentially, each block in $\mathcal{B}_i$ represents a non-idle period for the helper. In our example, there are two blocks: $\beta_1 = \{1,4,2,3\}$ and $\beta_2 = \{5\}$ with $s(\beta_1)=0, e(\beta_1)=8, s(\beta_2)=9,$ and $e(\beta_2)=10$. We can now focus on each block separately~\cite{baker1983preemptive}. For each block $\beta\in \mathcal{B}_i$, we find client $\ell\in \beta$ such that 
\begin{equation} \label{eq:alg2:ell}
   \ell := \argmin_{j\in \beta} \{e(\beta) +r_{j}^\prime \}.
\end{equation}
In our example, that would be client 4 for $\beta_1$ since $9 = \min \{8+5, 8+3, 8+8, 8+1\}$ and client 5 for $\beta_2$.
Then, we reschedule the tasks in $\beta$ such that the \texttt{bwd-prop} task of client $\ell$ is processed only during time intervals~(between $s(\beta)$ and $e(\beta)$) where no other client's task has arrived or is being processed. In our case, since $\beta_2$ contains a single client, no reschedulings are required within $\beta_2$, while, for $\beta_1$, client 2 ``moves'' to an earlier slot in the schedule (subject to its arrival time) and the task of client 4 is scheduled in slots where no other task is processed. This also decomposes the remaining set $\beta-\{\ell\}$ into a set $ \Gamma_{\beta}$ of subblocks~(according to the rule described above). In our example, $\Gamma_1=\{\beta_{11}, \beta_{12}\}$ as depicted in Fig.~\ref{fig:algo2}. Now,   for each subblock in $ \Gamma_{\beta}$, we find the client $\ell^\prime$ based on \eqref{eq:alg2:ell} and reschedule the tasks within this subblock (based on the same rules as above). In our example, $\beta_{11}$ needs no rescheduling, while, for $\beta_{12}$, $\ell^\prime$ is 2 since $10=\min\{7+3, 7+8\}$. The resulting schedule is optimal. In our case, client 3 will be processed upon arrival at the helper. The final optimal schedule has a makespan of 14, where client 3 will be the last one to finish the back-propagation of its part-1. This process runs in $\mathcal{O}(|\mathcal{J}_i|^2)$ time for helper $i$, so Algorithm 2 will run in $\mathcal{O}(\max_{i\in I}\{|\mathcal{J}_i|\}^2)$ time due to parallelization.

The ADMM-based solution method can be easily adapted in cases where clients own samples of different sizes. In such cases,  we can simply remove from the obtained schedules $\bm x^*$ and $\bm z^*$ the clients whose samples are completely processed (after a number of batch updates) and ``move'' the remaining clients earlier in the schedules~(subject to availability of their tasks at the helpers). Moreover, the assignments $\bm y^*$ do not need to change since helpers have already allocated memory for the model copies of the assigned clients. 

\section{Model Extensions \& A (Faster) Heuristic}
\label{sec:discussion_extentions} 
\textbf{Preemption Cost.} 
In certain systems (e.g., with very limited memory), preemption might induce further delays or costs (e.g., due to context switch~\cite{li2007quantifying}) 
that need to be taken into account while deciding on the schedule. This feature can be readily incorporated in our model without affecting the proposed solution method. 
In detail, let us denote with $\mu_i$ the switching cost that captures the delay induced at helper $i\in\mathcal I$ when switching between two tasks, i.e., $x_{ijt}\!=\!0$ and $x_{ij(t+1)}\!=\!1$, for some client $j\in\mathcal J$ and time interval $S_t$. Then, we can directly apply  the ADMM algorithm with the following modified constraint instead of~\eqref{eq:fwd_completion} in $\mathbb{P}_f$:
\begin{equation}\label{eq:completition-preemp}
\!	c_j^f\!=\! \phi_j^f \!+\! \sum_{i\in\mathcal{I}} l_{ij} y_{ij} \!+ \sum_{i\in\mathcal I} \sum_{t\in\mathcal T} \mu_i \big |x_{ijt} -x_{ij(t+1)}\big|,  \ \forall   j\in \mathcal{J}, \nonumber
\end{equation}
where the last term captures the cost of switching tasks when a preemption occurs and when a task has just started being processed. In a similar way, we can modify the constraints in \eqref{eq:completition} in $\mathbb P_b$ for variables $\bm z$, and use exact or inexact methods to solve the problems $\mathbb P_b^i$, as discussed in footnote \ref{foot_1}.

\textbf{A Scalable Heuristic (balanced-greedy).} 
Since the subproblems of Alg.~\ref{alg:fwd} are ILP problems, the ADMM-based method might induce high 
overhead. To exemplify, running the ADMM-based method on an ILP solver~(with exact solution) for a scenario of $J=70$ clients and $I=10$ helpers takes 14 min.  While such overhead may be tolerable in scenarios of this size, 
especially given the resulting time savings in total training makespan when compared to a baseline (see Sec.~\ref{sec:eval}), it might not be the case in larger problem instances. Moreover, this method decides on the assignments ($\bm y$) without taking into account the times $\bm p^\prime$ of \texttt{bwd-prop} tasks. Specifically, in our experiments, we noticed that when the processing times of \texttt{bwd-prop}, i.e., $\bm p^\prime$, are much larger than the times of \texttt{fwd-prop} tasks, i.e., $\bm p$,  long queues during \texttt{bwd-prop} may occur. This phenomenon can be alleviated by balancing the workload among the helpers.
We, thus,  propose a   heuristic that addresses these issues: it is of low complexity and \emph{balances} the client assignments among helpers in a greedy way.

 We propose \emph{balanced-greedy} that consists of two steps; it first decides on the client-helper assignments, and then on the scheduling. Specifically, it starts with $\bm x, \bm z, \bm y = \bm 0$ and:

1) The assignments follow a static load balancing algorithm~\cite{shah2017load}, where the load of helper $i$ is defined as the number of assigned clients, i.e., $G_i = \sum_j y_{ij}$.  For each client $j\in\mathcal{J}$, it finds the subset of helpers $Q_j$ with enough available memory  to allocate for $j$ (i.e., $Q_j := \{ i\in \mathcal{I} : m_i-\sum_h d_h y_{ih} \geq d_j\}$) and, based on $Q_j$,  it finds the helper $\eta$ with the least load, i.e., $\eta=\argmin_{i\in Q_j}\{G_i\}$. Balanced-greedy   assigns   client $j$ to $\eta$, i.e., $y_{\eta j} =1$, before proceeding to the next client.

2) The scheduling decisions $\bm x$ and $\bm z$ are made at each helper in a first-come-first-served~(FCFS) order~\cite{harchol2013performance}, i.e., for the \texttt{fwd-prop} tasks, the schedule $\bm x$  and the completion times $\bm c^f$  are  based on the release times $\bm r$, and, for the \texttt{bwd-prop} tasks,    $\bm z$  is based on   $\bm c^f + \bm l + \bm l^\prime$. In contrast to 
the ADMM-based method, balanced-greedy is non-preemptive. 

%% file: sections/evaluation.tex
\section{Performance Evaluation}
\label{sec:eval}

In this section, we evaluate the performance of our solution methods using measurements from our testbed's devices. 

\textbf{Dataset \& Models.} We use CIFAR-10~\cite{krizhevsky2009learning} and two NN models: \textit{(i)}~ResNet101~\cite{he2016deep}, and \textit{(ii)}~VGG19~\cite{simonyan2014very} for our training tasks. They are both deep convolutional NNs with $0.42$ and $2.4$ million parameters, and organized in $37$ and $25$ layers respectively. Hence, they may push resource-constrained devices to their limits when trained locally.

\textbf{Testbed.} The testbed's devices are listed in Table~\ref{tab:devices}, where the last two were employed as helpers. We also list the collected time measurements for a batch update for ResNet101 and VGG19. One of the devices (RPi~3) cannot fully train any of the two models locally due to its memory limitations. Furthermore, Jetson GPU's training times are comparable to the helpers' times, however, in practice,  the memory allocation for the GPU training can be very challenging~\cite{bai2022dnnabacus}.

\begin{table}[t] 
\caption{Testbed devices and average computing time (in sec) for a batch update, where the batch size is $128$ samples.}
\label{tab:devices}
\centering 
\begin{tabular}{lcc}
\hline
\textbf{Device} & \textbf{ResNet101}  & \textbf{VGG19} \\
\hline
 RPi~4 B Cortex-A72 (4 cores), 4GB & $91.9$ & $71.9$\\ 
 RPi~3 B+ Cortex-A5 (4 cores), 1GB & \multicolumn{2}{c}{not enough memory} \\ 
 NVIDIA Jetson Nano, 4 GB (CPU,GPU)& ($143$, $1.2$) & ($396$, $2.6$)\\ 
 VM 8-core virtual CPU, 16GB& $2$ & $3.6$\\ 
 Apple M1 8-core CPU, 16GB & $3.5$ &$3.6$\\ 
\hline
\end{tabular}
\vspace{-0.2cm}
\end{table}

\textbf{Setup.} In our simulations, the values of the input parameters of $\mathbb P$, i.e., $\bm r, \bm r^\prime, \bm p, \bm p^\prime, \bm l, \bm l^\prime$, are set according to the profiling data of the testbed (for the computation times) and findings on  Internet connectivity in France~\cite[p.56]{belson2017q4} (for the transmission times). We explore two scenarios that represent \emph{two levels of heterogeneity} in terms of devices, resources, and cut layers:

\noindent $\bullet$ \textbf{Scenario 1 (low heterogeneity):} Clients and helpers have the same parameters as in Table~\ref{tab:devices}, where the selection of the type of device for each client and helper is uniformly random.  Moreover, the entities' memory capacities are limited by their RAM size and all the clients' NNs have the same cut layers: layers $3$ and $33$ for Resnet101, and layers $3$ and $23$ for VGG19.  

\noindent $\bullet$ \textbf{Scenario 2 (high heterogeneity):}
To capture higher heterogeneity, the input parameters are devised by interpolating the time measurements of the profiled devices.  Also, the entities' memory capacities can vary from device to device, but upper-bounded by their RAM size.  In this scenario, 
 clients participate in the training with different cut layers~(randomly selected).

\begin{table}[t]
\caption{Suboptimality and speedup achieved by the ADMM-based method when compared to an ILP solver.}
    \centering
    \begin{tabular}{p{0.2cm}p{0.3cm}p{0.5cm}p{0.5cm}p{0.5cm}>{\centering\arraybackslash}p{2.3cm}>{\centering\arraybackslash}p{1.5cm}}
        \hline
        
         &  
          & \textbf{$J$} & \textbf{$I$} & \textbf{$T$} &  \textbf{suboptimality (\%)} & \textbf{speedup ($\times$)} \\ 
         \hline
        \multirow[c]{6}{*}{\rotatebox[origin=c]{90}{\small{Scenario1}}} & \multirow[c]{3}{*}{\rotatebox[origin=c]{90}{\tiny{Resnet101}}} &  \multirow[c]{2}{*}{$10$} & $2$ & $294$ & $0$ & $32.2$ \\ \cline{4-7} 
        &  &  & $5$ & $294$ & $2.6$ & $13.2$\\ \cline{3-7}
         &  & $15$ & $5$ & $399$ & $14.9$ & $37$ \\ 
        \cline{2-7}
         & \multirow[c]{3}{*}{\rotatebox[origin=c]{90}{\tiny{VGG19}}} &  \multirow[c]{2}{*}{$10$} & $2$ & $176$ & $0$  & $12.5$ \\ \cline{4-7} 
        &  &  & $5$ & $176$ & $0$ & $18.5$\\ \cline{3-7}
         &  & $15$ & $5$ & $211$  & $0$ & $22$ \\ \cline{4-7} 
        \hline
        \multirow[c]{6}{*}{\rotatebox[origin=c]{90}{\small{Scenario2}}} & \multirow[c]{3}{*}{\rotatebox[origin=c]{90}{{\tiny Resnet101}}} &  \multirow[c]{2}{*}{$10$} & $2$ &  $321$  & $8.2$ & $26.4$ \\ \cline{4-7} 
        &  &  & $5$ & $324$ & $10.2$ & $22.8$\\ \cline{3-7}
         &  & $15$ & $5$ & $445$ & $4.2$ & $17.2$\\
        \cline{2-7}
         & \multirow[c]{3}{*}{\rotatebox[origin=c]{90}{\tiny{VGG19}}} &  \multirow[c]{2}{*}{$10$} & $2$ &  $263$ & $3.2$ & $20.5$ \\ \cline{4-7} 
        &  &  & $5$ & $265$ & $0$ & $31.2$\\ \cline{3-7}
         &  & $15$ & $5$ & $331$ & $2$ & $52$ \\ 
       \hline
    \end{tabular}
    \vspace{-0.3cm}
    \label{tab:suboptimality}
\end{table}

\textbf{Results and Observations.} We proved that $\mathbb P$ is NP-hard, which led us to propose two scalable solution methods. Table~\ref{tab:suboptimality} shows the suboptimality and speedup achieved by the ADMM-based method when compared to Gurobi~\cite{gurobi}, one of the fastest ILP solvers~\cite{anand2017comparative},    that optimally solves $\mathbb P$. We observe that, in most cases, our method achieves less than $10.2\%$ suboptimality, with one corner case of $14.9\%$. However, even in this case, there is a $37\times$ speedup when compared to the solver. 
We highlight that these results derive from running less than $5$ iterations of Algorithm 1, while we may achieve smaller suboptimality 
with a larger number of iterations 
using techniques like the ones in~\cite{diamond2016general}.  Actually, ADMM may be tailored so that we can balance suboptimality and speed.

\theoremstyle{definition} \newtheorem{obs_optimal}[]{Observation} 
\begin{obs_optimal}
The ADMM-based method finds the optimal solution for $\mathbb{P}$ in several problem instances and achieves up to $52\times$ speedup when compared to an ILP solver.
\end{obs_optimal}

\begin{figure}[t]
\centering
\includegraphics[clip,width=0.4\textwidth]{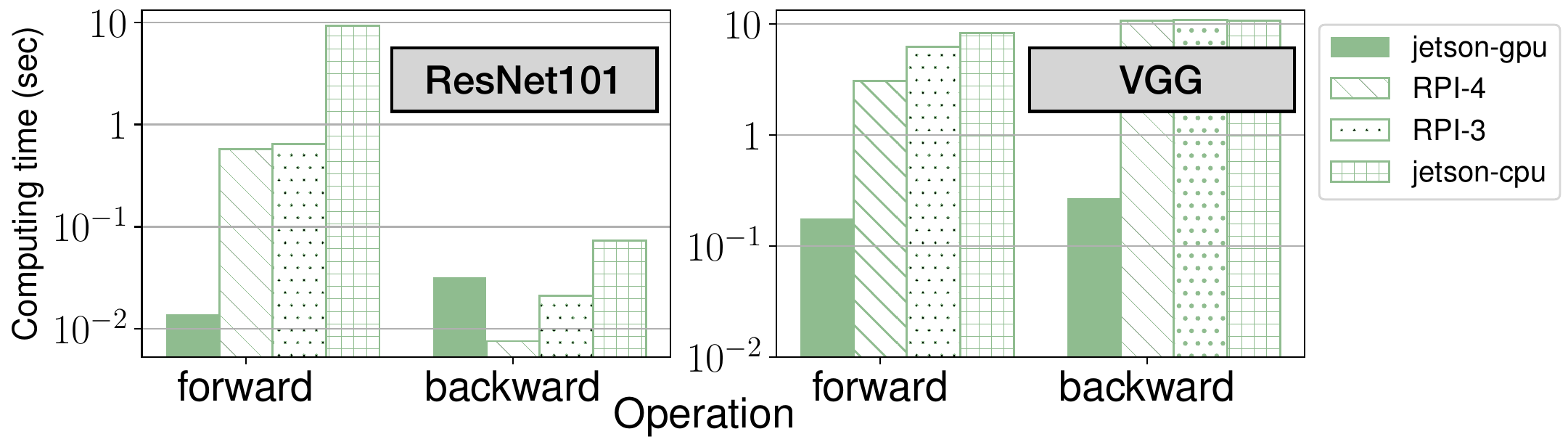} 
\caption{Profiled computing time (ms.) of part-1 for each device. \vspace{-0.2cm}
}
\label{fig:model_Part}
\end{figure}

We note that the numerical evaluations in Table~\ref{tab:suboptimality} were performed in small instances~(up to $15$ clients and $5$ helpers) that can be handled by ILP solvers. We observe that the lowest suboptimality gap is achieved for VGG19, which comes from the choice of cut layers (see above). In particular, Fig.~\ref{fig:model_Part} shows the processing times between forward and backward propagation per device for the two NNs. We see that these times can highly differ between forward and backward operation. Such asymmetries that can occur in SL further corroborate our approach towards jointly optimized assignments and scheduling.  

Next, in Table~\ref{tab:suboptimality}, we see that the time horizon ($T$) increases with the problem size, and our method achieves considerable speedups in execution time for very large $T$. In particular,  $T$ is directly related to the number of variables of the problem. This dependency can be critical   in cases where the input parameters $\bm r, \bm p, \bm l,$ etc. are in the order of thousands (e.g., when  in ms). For this reason, we explore next how tuning the time slot's length can affect the obtained solution (batch makespan).

\begin{figure}[t]
\vspace{-0.3cm}
	\centering  
\includegraphics[clip,width=0.4\textwidth]{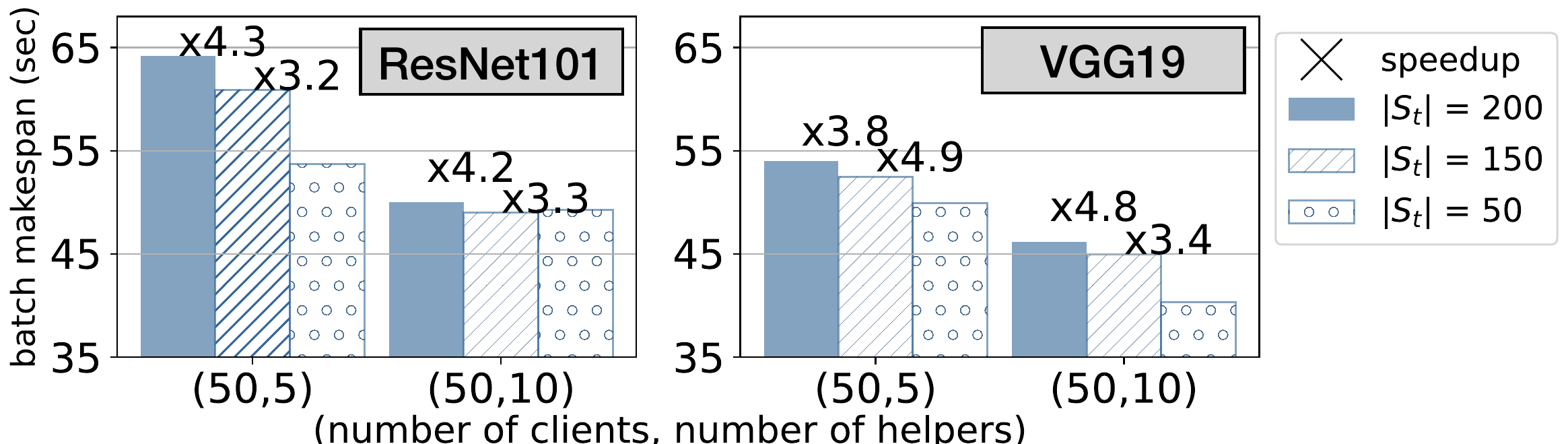} 
	\caption{Batch makespan obtained by the ADMM-based method for time slot length $|S_t|$ equal to $200$ ms, $150$ ms, and $50$ ms, in Scenario 1. The speedup is relative to the case $|S_t|=50$.\vspace{-0.3cm} }  
	\label{fig:granul_batch}
\end{figure}

In Sec.~\ref{sec:problem}, we discussed the impact of the time slot's length, i.e., $|S_t|$, on the frequency of preemptions. Furthermore,  as $|S_t|$ decreases,  the time horizon $T$ and the number of the problem's variables increase. To exemplify, a processing time of $400$ms would be translated into $2$,  $3$, or $8$ time slots when $|S_t| = 200$ ms, $|S_t| = 150$ ms, and $|S_t| = 50$ ms respectively. Since $T$ is defined based on the input processing and transmission times, its length will be the largest when $|S_t| = 50$. Moreover, in the case where $|S_t| = 150$, the processing time of our example is interpreted as $3$ slots, which
can overestimate the makespan. In fact, since the helper will need a bit less than $3$ slots to process the task, in a real-life implementation of the obtained schedule, it may be able to start processing the next task before the end of the 3rd slot. Therefore, in such cases,  the time slot's length may affect the accuracy of the obtained schedule. 

\theoremstyle{definition} \newtheorem{obs_granularity}[obs_optimal]{Observation} 
\begin{obs_granularity} 
As the length of the considered time slots $S_t$ increases, the obtained makespan increases, while the execution time decreases. This confirms an 
algorithmic tradeoff between the solution's precision and size of the solution space.
\end{obs_granularity}

Fig.~\ref{fig:granul_batch} depicts the makespan obtained by the ADMM-based method for 3 different $|S_t|$. The numbers on top of the bars are the speedups relative to the case   $|S_t|=50$, which has the highest overhead. We observe that the makespan is higher, in principle, as the $|S_t|$ increases. This is because large $|S_t|$ implies less frequent preemptions and a less precise solution.
Of course, as $|S_t|$ increases, the length of the time horizon $(T)$ decreases, which results in a speedup of up to $4.9\%$. Finally, we note that, for all the other experiments, 
we have used $|S_t|$ equal to $180$ms, for Resnet101, and to $550$ms, for VGG19.

\begin{figure}[t!]
\centering
    \centering
\includegraphics[clip,width=1\columnwidth]{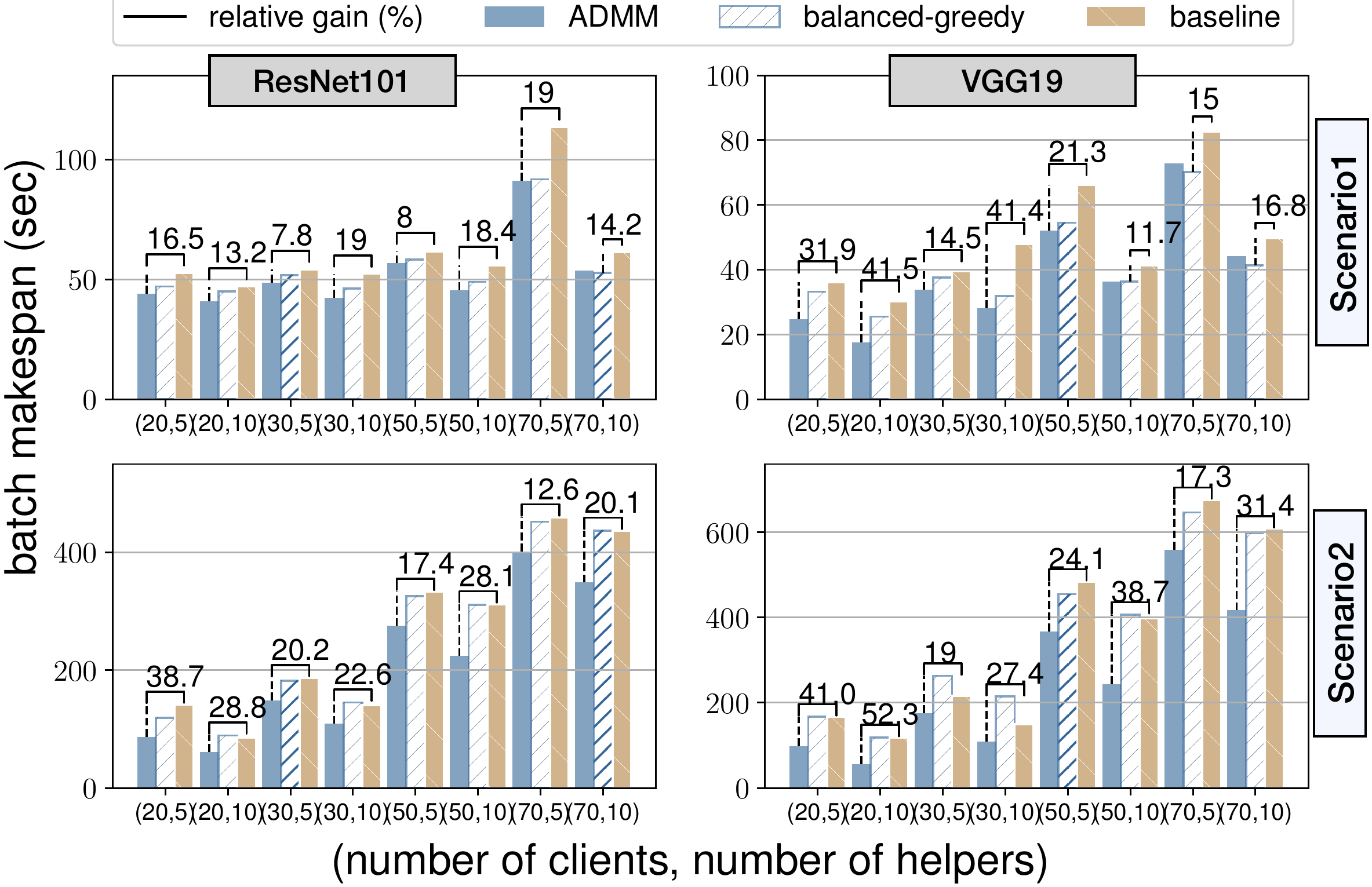}
    \caption{Comparison of the proposed methods with the baseline. \vspace{-0.2cm}
    }
    \label{fig:heur}
\end{figure}

Next, in Fig.~\ref{fig:heur}, we compare the proposed methods (ADMM-based and balanced-greedy) between them, and with a \emph{baseline scheme} that first decides on the client-helper assignments in a random way~(subject, of course, to memory constraints), and then schedules the tasks in a FCFS order. This baseline could be seen as a naive real-time implementation of parallel SL without proactive decisions on assignments or scheduling. 
We note that a straightforward comparison with related work (e.g., \cite{wang2023coopfl}) is not possible since, typically, its client-helper assignments are coupled with decisions on cut layers or other considerations, while it usually adopts a FCFS schedule.

We focus first on the comparison between the two proposed methods. We observe that the ADMM-based method finds a shorter makespan~(up to $32\%$) than the balanced-greedy in medium-sized scenarios (1 \& 2), i.e., up to 50 clients, and thus, it should be the preferred method.  However, as the number of clients grows and the queuing delays risk to increase, balancing the helpers' loads provides a better solution. 
Indeed, in Scenario 1~(top), the balanced-greedy achieves a better makespan than the ADMM-based method, making it, hence, the method of choice. 

Next,  as the heterogeneity of the network resources increases (Scenario 2, bottom Fig.~\ref{fig:heur}), the scheduling and assignment decisions become more crucial   for the makespan. Therefore, it is not surprising that the ADMM-based method outperforms the balanced-greedy 
by up to $48\%$, and thus, it should be the preferred method. We note that the problem instances of Scenario 2 contain a few helpers with very limited memory capacities that were not fully utilized by balanced-greedy. The presence of such helpers explains the fact that the batch makespan is larger in  Scenario 2 than in Scenario 1, since this implies long queuing delays in the other helpers. Nevertheless, these performance gains decrease as the number of clients increase, and given the discussion on the overhead of the two methods in Sec.~\ref{sec:discussion_extentions}, balanced-greedy should be preferred for very large scenarios (e.g., $\geq 100$ clients in our case) to avoid excess overhead.

\theoremstyle{definition} \newtheorem{obs_strategy}[obs_optimal]{Observation} 
\begin{obs_strategy} 
The numerical evaluations allow us to build a solution strategy based on the scenario's characteristics that achieves a shorter makespan than the baseline by up to 52.3\%.
\end{obs_strategy}

The observations above shape a \emph{solution strategy} that comprises the two proposed methods, and it is tailored based on the scenario at hand~(i.e., its heterogeneity and size). Also, we see that any improvement of one method over the other is larger in the VGG19. 
This reveals a  dependency on the NN, since NNs may differ on the cut layers, the size of the processing tasks~\cite{zhang2023privacy}, etc. We plan to further explore  this dependency and its implications on the makespan in future work.

Focusing now on the comparison of our strategy (i.e., ADMM-based or balanced-greedy depending on the scenario, as discussed above) to the baseline scheme, we observe that our proposed strategy consistently outperforms the baseline, achieving a shorter makespan. In detail, the baseline scheme decides on the assignments $\bm y$ without taking into account processing and transmission delays, which results in a larger makespan. Essentially, this confirms the need for workflow optimization in parallel SL. We notice that, in some instances where the preferred method is the ADMM-based one (e.g., (30,5) for VGG19), the baseline with random client-helper allocation may find a shorter makespan than balanced-greedy. This is because balanced-greedy may allocate clients to slower helpers, without taking into account that queuing delays might not be long in faster helpers for such medium-sized instances.

\begin{figure}[t]
	\centering  
\includegraphics[clip,width=0.5\textwidth]{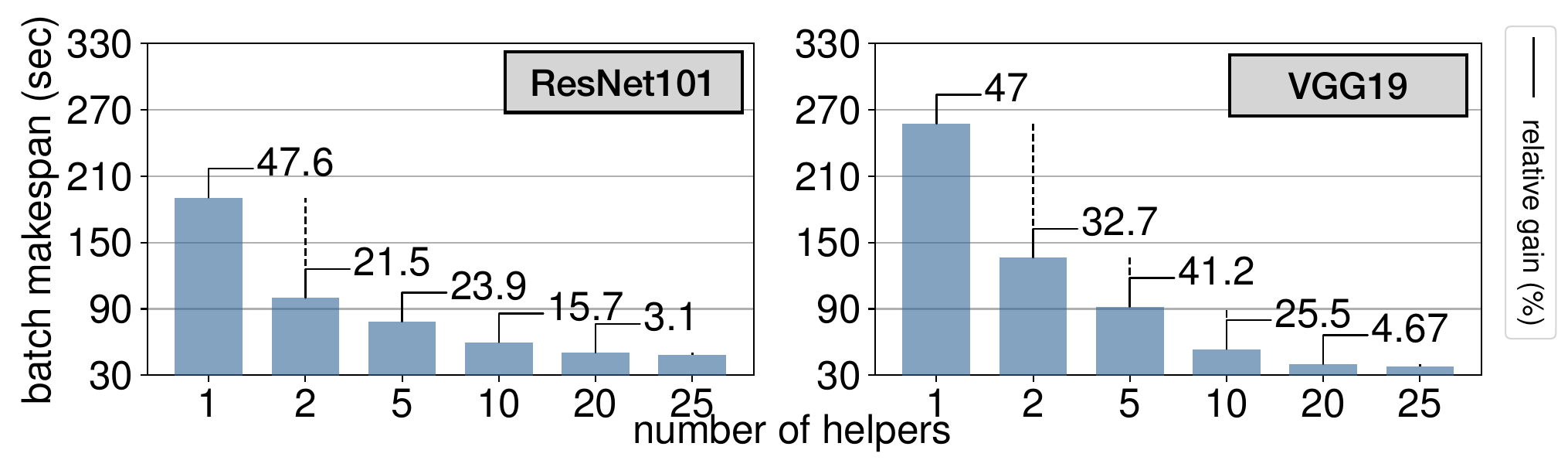} 
	\caption{Batch makespan obtained by the balanced-greedy method in Scenario 1 for $J= 100$ clients and varying $I$. \vspace{-0.3cm}}  
	\label{fig:dimish}
\end{figure}

Finally, in Fig.~\ref{fig:dimish}, we perform a sensitivity analysis with respect to the number of helpers in Scenario 1 where we depict the relative gains in batch makespan. Given the scenario's type and size, we employ balanced-greedy.

 \theoremstyle{definition} \newtheorem{obs_sens}[obs_optimal]{Observation} 
\begin{obs_sens} 
In a scenario of $100$ clients and $1$ helper, adding one more helper can dramatically decrease the batch makespan by  up to $47.6\%$.
\end{obs_sens}

Whereas, in the presence of $10$ helpers, the relative gains of adding more helpers are decreasing. Such observations provide useful insights for a real-life implementation of parallel SL and lead us towards future extensions of our approach where deployment or energy costs are included in our model. 

%% file: sections/conclusion.tex
\section{Conclusions and Future Work}
\label{sec:conclusion}

In this work, we formulated the joint problem of client-helper assignments and scheduling for parallel SL. We analyzed it both theoretically, proving it is NP-hard, and experimentally, using measurements from a realistic testbed.
We proposed two solution methods, one based on the decomposition of the problem, and the other characterized by a low computation overhead. Our performance evaluations led us to build a bespoke solution strategy comprising these methods that are chosen based on the scenario's characteristics. 
We showed that this strategy finds a  near-optimal makespan, while it can be tuned to balance suboptimality and speed. Also, it outperforms the baseline scheme by achieving a shorter makespan by up to $52.3\%$.  A natural direction for future work would be to decide on the NN's cut layers per client in conjunction with the proposed solution strategy.